# Yearly and seasonal variations of low albedo surfaces on Mars in the OMEGA/MEx dataset: Constraints on aerosols properties and dust deposits

Mathieu Vincendon[1], Yves Langevin[1], François Poulet[1], Antoine Pommerol[2], Michael Wolff[3], Jean-Pierre Bibring[1], Brigitte Gondet[1], and Denis Jouglet[1].

[1]*Institut d'Astrophysique Spatiale, CNRS/Université Paris Sud, Orsay, France.*

[2]*Laboratoire de Planétologie de Grenoble, UJF/CNRS, Bât. D de Physique, B.P. 53, 38041 Grenoble Cedex 9, France.*

[3]*Space Science Institute, 18970 Cavendish Road, Brookfield, WI 53045, USA.*

**Abstract:** The time variations of spectral properties of dark martian surface features are investigated using the OMEGA near-IR dataset. The analyzed period covers two Mars years, spanning from early 2004 to early 2008 (includes the 2007 global dust event). Radiative transfer modeling indicates that the apparent albedo variations of low to mid-latitude dark regions are consistent with those produced by the varying optical depth of atmospheric dust as measured simultaneously from the ground by the Mars Exploration Rovers. We observe only a few significant albedo changes that can be attributed to surface phenomena. They are small-scaled and located at the boundaries between bright and dark regions. We then investigate the variations of the mean particle size of aerosols using the evolution of the observed dark region spectra between 1 and 2.5 µm. Overall, we find that the observed changes in the spectral slope are consistent with a mean particle size of aerosols varying with time between 1 and 2 µm. Observations with different solar zenith angles make it possible to characterize the aerosol layer at different altitudes, revealing a decrease of the particle size of aerosols as altitude increases.

1.  **Introduction**

The atmosphere of Mars constantly contains suspended dust particles that interact in a complex manner with the surface (Sagan et al., 1972). These aerosols are associated with dust deposits on bright regions. Therefore, the settling of aerosols after major storms is expected to increase the albedo of dark regions (Balme et al., 2003; Kinch et al., 2007). The increasing contrast of the main low albedo features in the first few months after the storm season has been attributed to an efficient clean-up process over these dark regions (Sagan et al., 1972; Pleskot and Miner, 1981; Greeley et al., 2005; Rogers et al., 2007). Competitive deposition and clean-up is expected at boundaries between dark and bright regions (Geissler et al., 2008), which results in slow changes of the shape and extent of dark regions (Sagan et al., 1972; Lee, 1986; Christensen, 1988; Geissler 2005; Szwast et al., 2006). Such long term changes have an impact on the overall radiative balance, and hence, the climate of Mars (Cantor et al., 2007; Fenton et al., 2007).

Albedo maps provided by IRTM onboard Viking for a period covering a complete Martian year during which a global dust storm occurred have been analyzed by Pleskot and Miner (1981), who concluded that most large-scale variations of surface albedos seen from orbit can be attributed to variations of the atmospheric dust content. Mallama (2007) also observed that apparent albedo variations are strongly influenced by aerosol variations. These variations of the reflectance are strongly dependent on particle size in the near-infrared (Drossart et al., 1991). Clancy et al. 2003 defined three regimes with different mean grains size from TES data at low to mid-latitudes (45°S – 45°N) depending on latitude, solar longitude and optical depth. In a companion paper, Wolff and Clancy (2003) showed a distinct increase in the mean particle size associated with the onset of the 2001 global dust event. More recently, Wolff et al. (2006) and Soderblom et al. (2008) observed variations of the mean particle size at the MER landing sites, higher mean particle size being observed during periods of higher opacities.

In this paper, we investigate the variations with time of the apparent surface reflectance over low albedo regions in the low to mid-latitudes using OMEGA/MEx observations. The OMEGA dataset consists of hyperspectral images with a spatial resolution ranging from 300m to 5km, with a wavelength range from 0.3 µm to 5.1 µm. OMEGA has been operating since early 2004, providing more than two Martian years of data, including the 2007 high opacity dust storms. Most of the planet has been observed at a low spatial resolution (km scale), with frequent overlaps. These observations confirmed the major role of aerosols in the near-IR spectral range (Drossart et al., 2005, Vincendon et al., 2007a). Therefore, one must explicitly consider the effects of aerosols when interpreting observations in this spectral range in terms of surface characteristics.

In our analysis, we first construct time series of OMEGA observations of different dark regions of Mars at low to mid-latitudes (section 2). Next, we modeled the observed variations of the reflectance with time and/or photometric angles using a multiple-scattering radiative transfer code (section 3). Simultaneous PanCam/MER measurements of the optical depth are considered for comparison. Results are then discussed in terms of surface dust deposit occurrences and aerosols size variations (section 4).

## 2. Observations

### 2.1. The OMEGA dataset

The *Observatoire pour la Minéralogie, l'Eau, les Glaces et l'Activité* (Bibring et al., 2004) has observed almost the entire planet with a resolution of 2 to 5 km, beginning in January 2004. High-resolution observations (≥300m), performed during the lowest altitude portions of the Mars Express orbit, cover almost 10% of the surface. The number of overlapping tracks increases at high latitudes with the nearly polar orbit of Mars Express. Most terrains have been observed at least two times, with a set of locations observed as many as ten times. Most observations of mid-latitude dark terrains (40°S – 40°N) are obtained with the nominal nadir-pointing mode. In this configuration, the phase angle is equal to the solar zenith angle, which varies in a complex manner depending upon the latitude, the season and the local time of the observation. When the 3-axis drift mode is implemented for high altitudes of Mars Express (>4500km), OMEGA performed observations with emergence angles (also known as emission angles) that significantly depart from nadir. Additional non-nadir observations of the Martian surface have also been obtained during limb scans.

The OMEGA wavelength range (0.3μm – 5.1μm) corresponds to the maximum of light scattered by dust aerosols on Mars (e.g. Erard et al., 2001), which also includes several diagnostic features of water ice (1.5μm, 2μm, 3μm). OMEGA is therefore ideally suited for the study of Martian aerosols.

The reflectance of dark surfaces at a given wavelength $\lambda$ is defined by the "reflectance factor" (hereafter noted RF($\lambda$)) defined by I($\lambda$)/(Fcos(i)); where *i* is the solar incidence angle, or solar zenith angle, measured from the normal to the reference ellipsoid). RF($\lambda$) corresponds to the albedo for a lambertian surface. All observations of a same terrain should have the same RF if: the surface is lambertian, the surface properties do not change with time, and the atmosphere is free of aerosols.

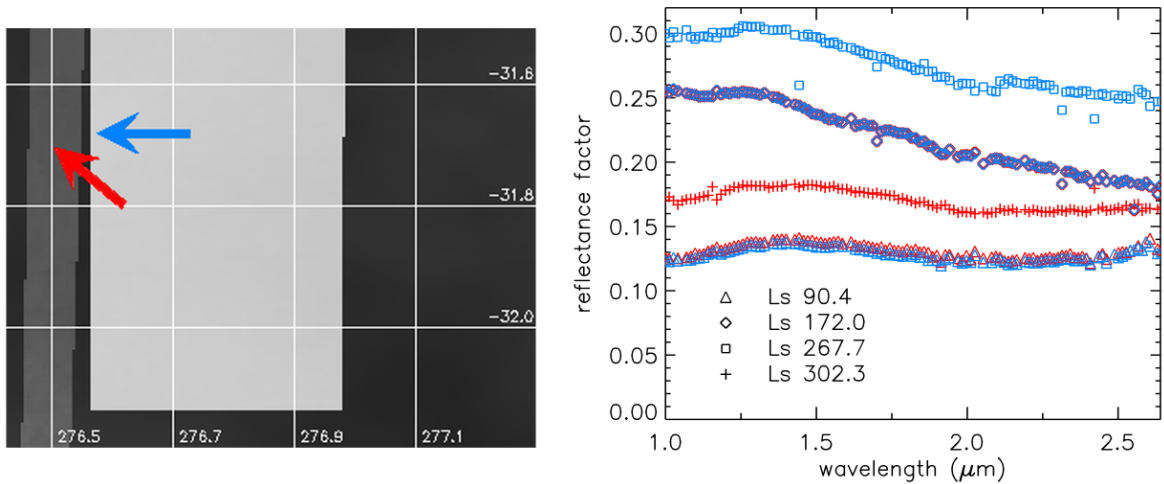

**Figure 1:** *(Left) Albedo map constructed with 3 observations obtained at different solar longitudes. 2 narrow tracks (dark and light gray, $L_S$ 302.3° and Ls 267.7° respectively) overlap a wide track (black, Ls 90.4°). Another wide track has been obtained at $L_S$ 172°. Two nearby regions (red and blue arrows) have been selected within the narrow tracks. (Right) Spectra of the two selected regions for different values of $L_S$. They are similar for the two wide tracks which cover both regions (triangles and diamonds), demonstrating that the reflectance properties of the red and blue areas are similar. It is therefore possible to include spectra of these two areas obtained at different values of $L_S$ (blue squares and red crosses) in a single time series from $L_S$ 90.4° to $L_S$ 302.3°.*

## 2.2. Constraints for selecting observations

Selected regions of interest correspond to surface albedo at 1 µm of 15% or less. The selection of the spectra constituting our time series for a given dark region requires the careful consideration of the following points:

- We have restricted our analysis to regions for which at least one observation is available at a time when the contribution of aerosols as observed by the Mars Exploration Rovers is low (effective optical depth – $\tau/\cos(i)$ – less than 0.6). This observation will be used to estimate of the actual dependence of the reflectance of the area with wavelength.

- The contribution of aerosols varies with the incidence angle measured from the normal to the reference ellipsoid, whereas the contribution of the surface is linked with the local incidence angle, which is dependent on local slopes. We therefore need to constrain our analysis to relatively flat regions. This is checked with the MOLA elevation measurements (Smith et al., 1999), which does not preclude local slopes or shadows at smaller scales.

- High-resolution tracks (width of about 5 – 20 km) can cover homogeneous terrains at slightly different locations (a few kilometers apart). In order to increase the time resolution of our sequences, we have included non-overlapping observations after checking that reflectance properties where similar to the nominal region, as illustrated in *Figure 1*.

- Observations with significant contributions from ice-rich aerosols are not included in the time series. They can be detected from the presence of weak absorption bands at 1.5 µm and 2 µm, and a stronger absorption band at 3 µm (Langevin et al., 2007), as shown in *Figure 3*. Fortunately, water ice aerosols are not detected frequently in OMEGA observations as most dark regions are found south of the aphelion cloud belt (*Figure 2*).

- The nadir observations take into account photons coming from the surface within the IFOV, and a small fraction of photons coming from nearby areas that are then scattered by aerosols into the IFOV (Langevin et al., 2008). Monte-Carlo simulations using the model of Vincendon et al. (2007a) demonstrates that this contribution becomes negligible at distances of 25 km or more from the IFOV. We have therefore selected regions of interests with variations of the albedo lower than 0.01 over a circle 50 km in diameter.

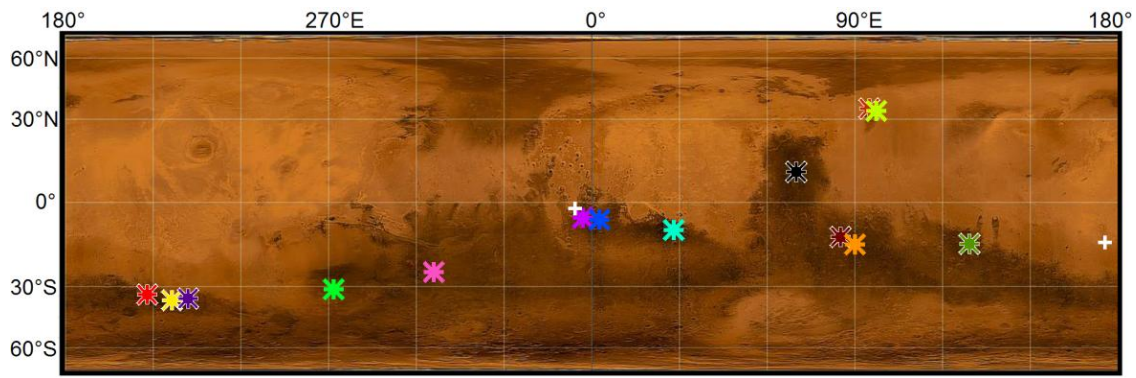

**Figure 2:** *Locations of the selected time series on a background of a cloud-free mosaic constructed from MGS images obtained during the period 1999 to 2001 (Geissler, 2005, modified by Szwast et al., 2006). The two white crosses indicate the position of the MERs Opportunity (near 0°) and Spirit (near 180°).*

### 2.3. Selected time series of observations

The locations of the 14 time series defined from the selection procedure presented in section 2.2 are indicated on *Figure 2*.

An example of a time series (black stars on the map) showing a wide variety of reflectance spectra is presented in *Figure 3* for the [0.3 µm – 4 µm] wavelength range. In general, we observe significant variation in spectral slopes between 1 µm and 2.5 µm (from nearly flat spectra to spectral slope of 40%) and albedo levels (from 0.1 to 0.3 in the red / 1 µm range). Observations with steep blue spectral slopes and strong increase of the reflectance factor correspond to those that include a large contribution of scattering by aerosols (dust storm or high solar zenith angles). In the next section, we present modeling results, which aim to determine whether part or all of the observed spectral variations can be attributed to aerosol effects. We focus on the near IR range (0.9 µm to 2.6 µm) which corresponds to one of the three detectors of OMEGA (the C-channel). This spectral range provides typically the strongest constraints on the size of aerosols (Clancy et al., 2003); it can be compared to visible studies as the effect of aerosols varies weakly between 0.7 and 0.9 µm (Ockert-Bell et al., 1997; Tomasko et al., 1999) and it is free of thermal contribution. Also of note is the fact that the C-channel is the detector with the lowest radiometric or geometric uncertainties (Bellucci et al., 2006; Jouglet et al., 2008), making comparisons more precise.

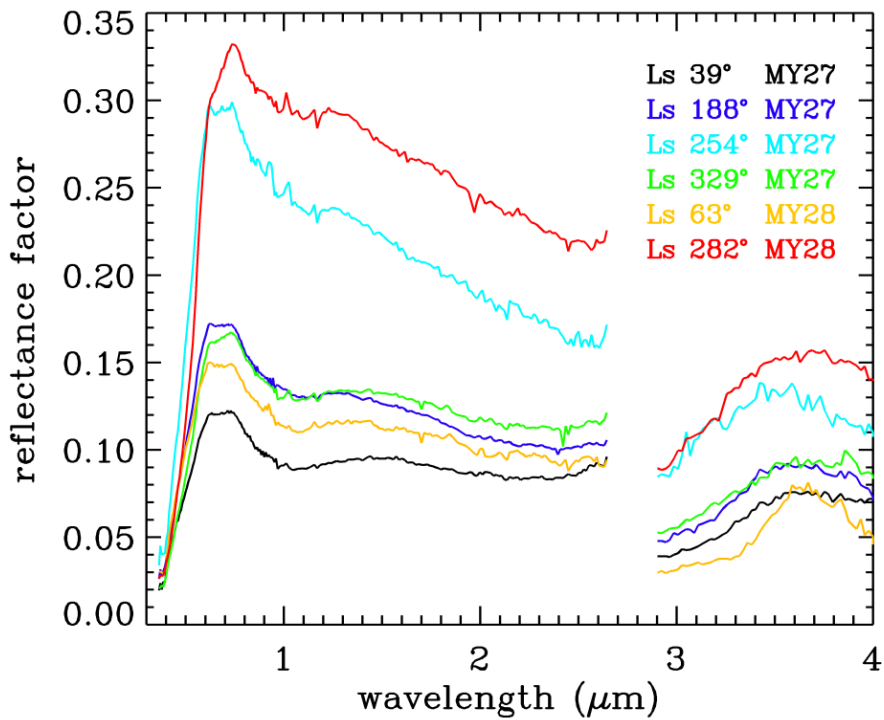

**Figure 3:** *Time series of observed reflectance spectra for the region at 70.1° E, 13.2° N (Syrtis Major). Solar longitudes and martian years (using the numbering of Clancy et al., 2000) are indicated. Wavelengths >2.9 µm are processed using the approach of Jouglet et al. (2008). A wide diversity of spectra is observed. Black: low atmospheric contribution.*

*Yellow: the absorption bands at 1.5, 2 µm and the shape of the 3 µm band are indicative of water ice clouds (Langevin et al., 2007). Light blue: a strong decreasing slope is observed at high solar zenith angle (80°). Red: July 2007 dust storms. Dark blue and green: significant differences in spectral slope (15 and 30%) are observed even for spectra with similar albedo levels.*

### 3. Modeling assumptions

In this section, we present models of the observed spectra under the assumption that the surface reflectance does not change with time. The results in terms of the derived aerosol parameters will be compared to the ground truth provided by MER/PanCam measurements (Lemmon et al., 2004) and discussed in section 4. We model Martian spectra observed by OMEGA using the Monte-Carlo based approach developed by Vincendon et al. (2007a, 2008). This type of algorithm simulates the path of photons through a layer containing aerosols, including interactions with the surface. Multiple scattering is handled explicitly, with no simplifying assumptions. The geometry of the model is one-dimensional plane-parallel. The surface reflectance is characterized by a prescribed photometric function and albedo. Three parameters describe the optical behavior of aerosols: the single scattering albedo, the single scattering phase function and the total optical depth. These five parameters are a priori wavelength dependent. Our approach involves the use of look-up tables of apparent reflectance factors, built by simulating the path of a large number of photons (typically $10^6$) for each set of input parameters and observing geometries. The removal of gas bands is performed using the "OMEGA volcano-scan approach" described in Langevin et al. (2007). In this method, the observed spectra are divided by a typical atmospheric spectrum scaled to the path length of photons in the atmosphere for each given observation.

### 3.1. Surface photometric function

Because our observations for a given location were obtained under different illumination conditions, it is useful to investigate the expected variations of the surface contribution with photometric angles. Johnson et al. (2006a, 2006b) have extensively studied the photometric properties of different surface textures observed at both MER landing sites. Observed surface behaviors in the visible wavelengths depart from the Lambert case by at most ±10% in the [25° – 90°] range of phase angles, the surface reflectance generally increasing in the anti-solar direction. For the data considered here, the variations of the phase angle are due generally to changes in the solar zenith angle. Given the limited range of the photometric angles in our dataset, it is difficult to generalize the results of Johnson et al. for our modeling efforts. However, recent laboratory measurements provide some guidance. Pommerol and Schmitt (2008) designed a set of laboratory experiments to study the photometric properties of the Martian surface in the near-IR wavelengths, using a spectro-gonio-radiometer (Brissaud et al., 2004). We have extracted from their measurements those corresponding to the geometry of most of our observations: normal emergence with various incidence angles. In addition, we have obtained measurements of two new surface analogues specifically for this study (*Figure 4*). The photometric properties of the selected analogues do not vary significantly with wavelength in the 1µm – 2.5µm range. Variations of the RF with incidence (±4% in the 25° - 70° range for the surface analogues considered here) are at least one order of magnitude lower than the observed variations in our time series. Therefore, a Lambertian behavior represents a good first approximation for the martian surface as observed with varying solar incidence angle. Additional justification for this assumption may be seen in the fact that the photons reaching a martian surface element sample a wide range of effective incidence angles (Clancy and Lee, 1991). This perturbation from the canonical solar incidence angle is due to aerosol scattering before photons reach the surface. We have characterized this distribution of arrival angles at the surface through our Monte-Carlo code (*Figure 5*). In effect, 40% of photons reaching the surface have been scattered outside the original solar direction for a solar incidence angle $i$ of 10° and a moderate optical depth of 0.6. This fraction is larger than 70% at $i$=70°. Consequently, the behavior of the Martian surface as a function of the solar zenith angle is even closer to a Lambert surface than estimated from measurements at well-defined incidence angles.

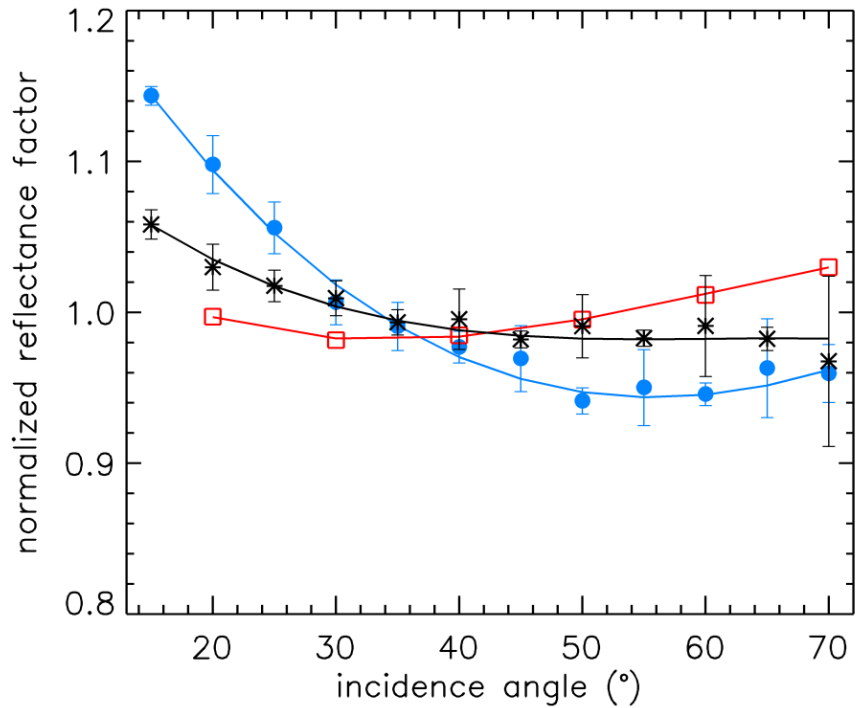

**Figure 4:** *Laboratory measurements of the reflectance factor in the near-IR (λ ~ 1 µm) for a nadir viewing geometry and different incidence angles (Pommerol and Schmitt, 2008). Red squares (λ = 1 µm): small-grained volcanic tuff (radius: 25– 50 µm). Black crosses (λ = 1.3 µm): small-grained volcanic tuff (radius: 25–50 µm) covered by mm size grains of the same material (surface coverage by large grains: 30–35%; grains are partially buried). Filled blue circles (λ = 1 µm): mixture of sandsized grains (radius ranging from 140 to >600 µm) of a dark basalt (in agreement with thermal inertia measurements of dark regions, see e.g. Palluconi and Kieffer, 1981). Each point is the average of 20 measurements, with standard deviations indicated as "error bars." Lines correspond to a third order polynomial fit of each analogue.*

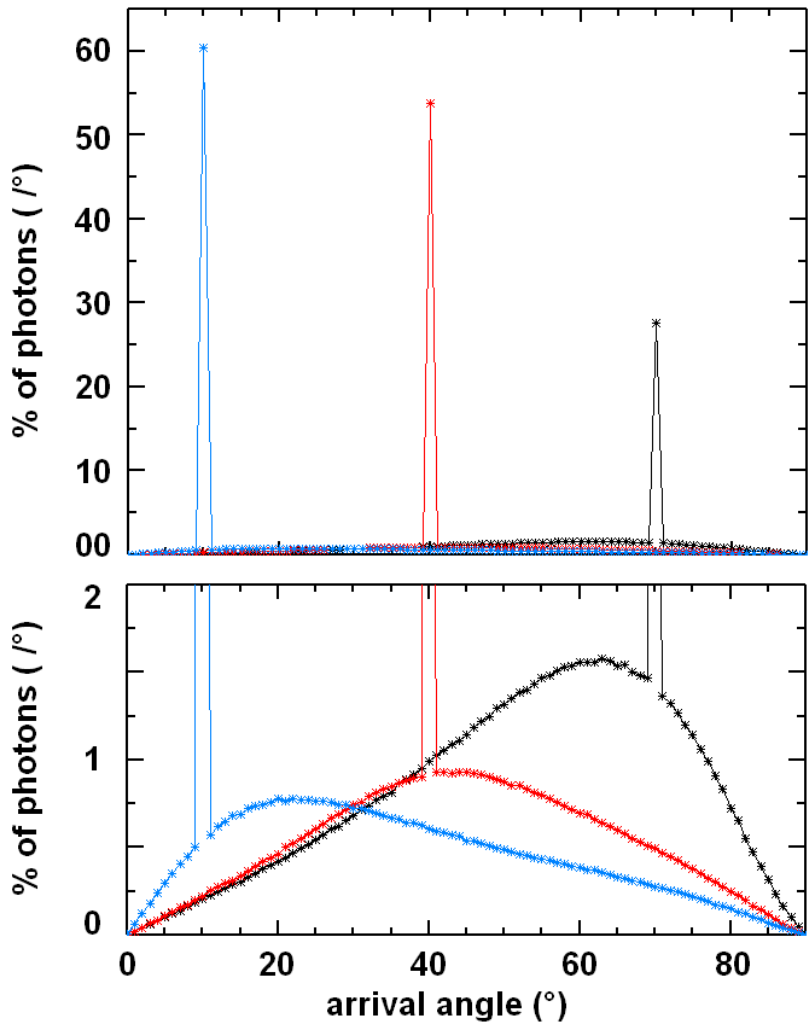

**Figure 5:** *Distribution of photons that reach the surface as a function of the arrival angles (0° is zenith). Simulations are performed using the model and aerosols parameters described by Vincendon et al. (2007a, 2007b). The aerosols optical depth is 0.6. Three different solar incidence angles are shown (i = 10°, blue; i = 40°, red; and i = 70°, black).*

### 3.2. Assumptions on the optical properties of aerosols

We observe variations of the spectral slope of dark regions with time and photometric angles (*Figure 3*). If the surface does not change, this requires variations of the aerosol properties. Solar longitudes and solar zenith angle vary simultaneously in our sequences. In *Figure 6*, we evaluate the amplitude of the spectral slope variations with solar zenith angle due to expected variations of the phase function of aerosols with wavelength. We computed two wavelength dependent phase functions using the code of Mishchenko and Travis (1998). We considered two potential Martian dust aerosol size distributions: 1 – a bimodal distribution of spherical particles that combine large aerosols generally seen during storms (Toon et al., 1977; Clancy et al., 2003) and a smaller particle size generally seen as the altitude increases (Chassefière et al., 1992; Montmessin et al., 2006; Rannou et al., 2006); 2

– a size distribution with a mean particle size of 1.4 μm of randomly oriented oblate cylinders with a 2:1 axial ratio (Wolff et al., 2006). Indices of refraction are taken from Ockert-Bell et al. (1997). Weak variations of the phase function with wavelength are inferred from these models. This results in a ratio of light scattered by aerosols at 1 μm to light scattered at 2.6 μm that ranges between 1.1 and 1.3 (*Figure 6*), while this ratio varies between 1 and 4 for our spectra. This suggests that observed variations of the spectral slope may not be predominantly due to variations of the phase function with wavelength. As a result, for simplicity, we will adopt a wavelength independent phase function. Two phase functions retrieved on Mars in the near-IR by other authors will be tested: a Henyey-Greenstein function with an asymmetry parameter of 0.63 (retrieved by Ockert-Bell et al. 1997 between 1 μm and 2.5 μm) and the phase function derived by Tomasko et al. (1999) at 0.965 μm. We have selected a constant single scattering albedo of 0.97 according to Vincendon et al., 2007a. This value has been recently confirmed by other authors using different approach or dataset (Wolff et al., 2007; Määttänen et al., 2009). The last aerosol parameter to consider is the extinction cross section, being proportional to the optical depth. We treat it as a free, wavelength dependent parameter, as expected from size variations of aerosols (see e.g. Clancy et al., 2003, figure 13).

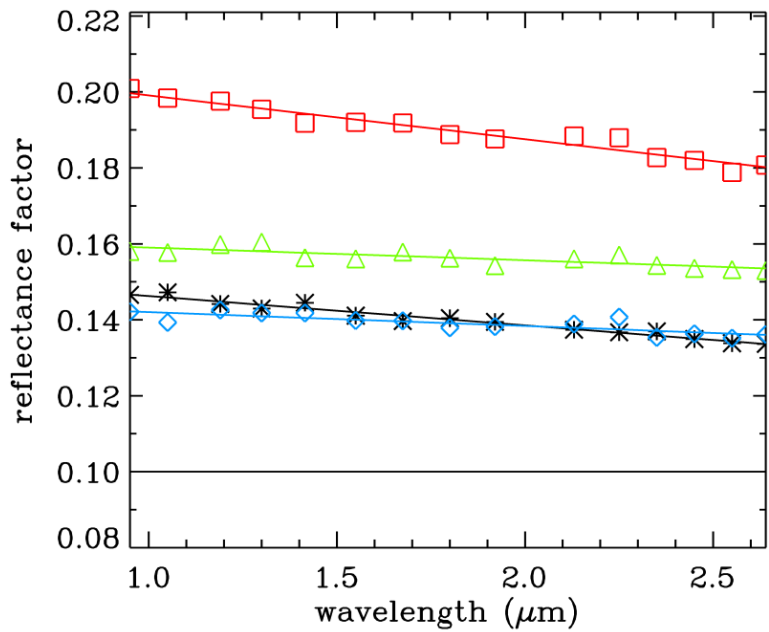

**Figure 6:** *Modeled near-IR reflectance spectra as seen above a dark surface (albedo of 0.10, solid black line) through an aerosols layer (constant optical depth of 0.8) with the wavelength dependent phase function computed with the monomodal size distribution described in the text. The viewing geometry is nadir with different solar incidence angles, i, shown. Red squares: i = 80°; green triangles: i = 60°; blue diamonds: i = 40°; black stars: i = 10°. Modeled RF are calculated for 15 wavelengths (symbols); Smooth interpolated lines are indicated. A wavelength dependent phase function can create a varying decreasing slope in the light backscattered by aerosols. However, the amplitudes of resulting variations are small compared to observations.*

## 4. Interpretation of the observed spectra in terms of aerosol properties and surface variations.

For each time series, we estimate the surface albedo spectrum from the observation with the lowest aerosol contribution. These observations are obtained during clear atmospheric conditions ($L_S \sim 30° - 130°$). Even in this case, we performed a small aerosol correction to the data. The low to mid-latitude optical depth is relatively homogeneous in that $L_S$ range (Smith, 2004). We can therefore use the optical depth determined at 0.9 µm by MER/PanCam measurements (Lemmon et al., 2004, 2006) at the time of the observation. Optical depths at other wavelengths are derived using the "typical" aerosol wavelength dependent cross section retrieved previously from other locations with OMEGA (the optical depth monotonically decreases between 1µm and 2.65µm by a factor of 1.9; Vincendon et al., 2007b, 2008). Uncertainties resulting from these approximations will be discussed in section 4.1. With the resulting surface spectrum, we constructed the model spectra that provided the best fit of each observation by adjusting the optical depth at 0.9 µm as well as the wavelength dependence of the optical depth, where this latter aspect was done through linear combinations of optical depth functions computed with Mie theory for different grain sizes (Clancy et al., 2003, figure 13). We characterize the result of each model fit using two parameters: the optical depth at 0.9 µm $\tau(0.9\mu m)$, and the "slope factor" $S$ defined by the ratio between the optical depth at 1 µm and that at 2.5 µm.

For illustration purposes, two time series are presented in *Figure 7* and *Figure 8*, with the associated retrieved aerosol parameters listed in *Table 1* and *Table 2*. For these two dark areas, the evolution of the spectral properties with time can be well modeled through variations in the optical thickness of aerosols and the assumption of an unchanging surface reflectance.

### 4.1. Uncertainties

Relative errors on the retrieved parameters are mainly related to the assumptions made when deriving the surface albedo spectrum of each time series. Uncertainties on the optical depths used for the correction (typically $\pm 0.2$) result in uncertainties between $\pm 5\%$ and $\pm 10\%$ for each surface spectrum. Errors due to photometric effects are estimated to be $\pm 2\%$ for most observations (see section 3.1), which do not greatly influence the total uncertainties when combined quadratically. The way these errors propagate to the retrieved parameters $\tau$ and S are indicated on *Figure 9*, *Figure 13* and *Figure 15*. The signal to noise ratio of OMEGA is greater than 100 for most observations (Bibring et al., 2005), which means that corresponding error bars can be neglected.

Systematic errors result on the one hand from calibration uncertainties of OMEGA data and on the other hand from assumptions about the optical parameters of dust aerosols. OMEGA absolute uncertainties are estimated to be $\pm 10\%$. A shift of $\pm 10\%$ on both surface and observed spectra results in optical depth uncertainties of $\pm 10\%$ for moderate aerosols loading ($\tau(1\mu m) = 0.5$, i = 45°) and of $\pm 20\%$ for high aerosols contribution ($\tau(1\mu m) = 1$, i = 75°). The uncertainty on the single scattering albedo is low (see section 3.2), and the impact of changing the phase function is illustrated in *Figure 10*.

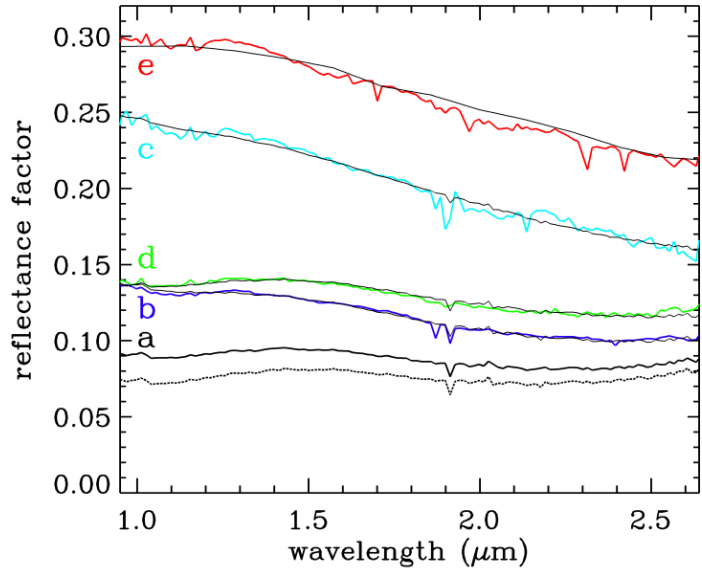

**Figure 7:** *Time series of observed spectra (thick solid color lines) for a region at 70° E, 13° N (Syrtis Major). The reflectance spectrum of the surface derived from observations with a low optical thickness is indicated as a dotted line. Best fit models for the observations are shown as thin solid lines. The solar longitude, the photometric angles and the results of the model fit are indicated in Table 1.*

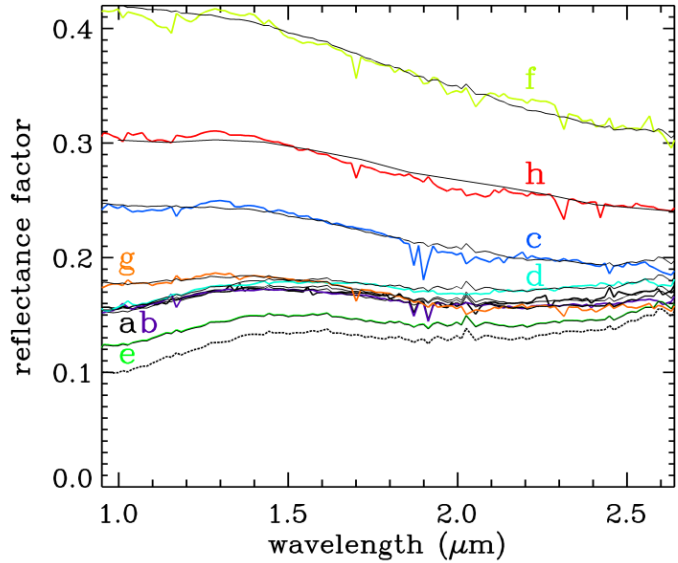

**Figure 8:** *Time series of observed spectra (thick solid color lines) for a region at 208° E, 34° S (Sirenium). The reflectance spectrum of the surface derived from observations with a low optical thickness is indicated as a dotted line. Best fit models for the observations are shown as thin solid lines. The solar longitude, photometric angles and results of the model fit are indicated in Table 2.*

**Table 1:** *Season, photometric angles and results of the model for observations of Figure 7. The optical depth τ (0.9 μm) and the slope factor S are assumptions for the observation with the lowest optical thickness (black in Figure 7).*

|   | Color | Solar Longitude | Mars Year | Incidence | Emergence | Phase | τ(0.9μm) | S ("Slope factor") | RMS (%) |
|---|---|---|---|---|---|---|---|---|---|
| a | black | 39° | 27 | 38° | 2° | 36° | 0.35 | 1.8 | - |
| b | dark blue | 188° | 27 | 42° | 43° | 38° | 0.69 | 2.1 | 0.02 |
| c | sky blue | 254° | 27 | 80° | 4° | 83° | 1.08 | 3.3 | 0.06 |
| d | green | 329° | 27 | 30° | 5° | 32° | 0.99 | 1.4 | 0.03 |
| e | red | 282° | 28 | 40° | 1° | 40° | 2.9 | 1.6 | 0.07 |

**Table 2:** *Season, photometric angles and results of the model for observations of Figure 8. The optical depth τ (0.9 μm) and the slope factor S are assumptions for the observation with the lowest optical thickness (green in Figure 8).*

|   | Color | Solar Longitude | Mars Year | Incidence | Emergence | Phase | τ(0.9μm) | S ("Slope factor") | RMS (%) |
|---|---|---|---|---|---|---|---|---|---|
| a | black | 152.3° | 27 | 46° | 4° | 45° | 0.78 | 1.5 | 0.05 |
| b | purple | 187.0° | 27 | 39° | 9° | 45° | 0.87 | 1.8 | 0.04 |
| c | blue | 243.0° | 27 | 72° | 2° | 73° | 1.12 | 2.6 | 0.07 |
| d | sky blue | 314.4° | 27 | 16° | 10° | 19° | 1.09 | 1.2 | 0.04 |
| e | green | 92.8° | 28 | 63° | 2° | 63° | 0.25 | 1.8 | - |
| f | light green | 176.7° | 28 | 66° | 56° | 93° | 0.95 | 2.1 | 0.18 |
| g | orange | 213.7° | 28 | 61° | 1° | 60° | 0.75 | 2.6 | 0.12 |
| h | red | 273.0° | 28 | 10° | 1° | 11° | 3.38 | 1.6 | 0.13 |

### 4.2. Comparison with observations by the Mars Exploration Rovers (MER)

In *Figure 9* and *Figure 10*, we compare our retrieved optical depths with nearly-simultaneous measurements by the MERs (Lemmon et al., 2004, 2006 and personal communication). We use the Ockert-Bell et al. (1997) phase function in *Figure* 9 and *Figure* 10b and the Tomasko et al. (1999) phase function in *Figure* 10a. The elevation varies between - 4.5 km and 3 km for the different regions analyzed here. Generally, one considers that optical depth varies with elevation (i.e., surface pressure) when dust is well-mixed in the atmosphere (Zazova et al., 2005). However, it is important to note that the optical depths are often lower at the Spirit landing site compared to those for Opportunity, despite the slightly lower altitude (-1.9km vs. -1.5km). Thus, the altitude does not exclusively control the mean optical depth. Nevertheless, we have scaled all the retrieved optical depths to the elevation of the rover in *Figure* 10b. The variations of the modeled optical depth in our data are consistent with those measured by the MERs, regardless of the specific phase function adopted (*Figure 9* and *10a*) and of the assumption of elevation variations (*Figure 9* and *Figure 10b*). When combined with the evolution of the nominal spectral fits (under the assumption of a single surface reflectance function for each location), the correspondence of the retrieved optical depths of aerosols and the MER measurements strongly suggests that observed spectral variation from observation-to-observation is dominated by aerosol effects.

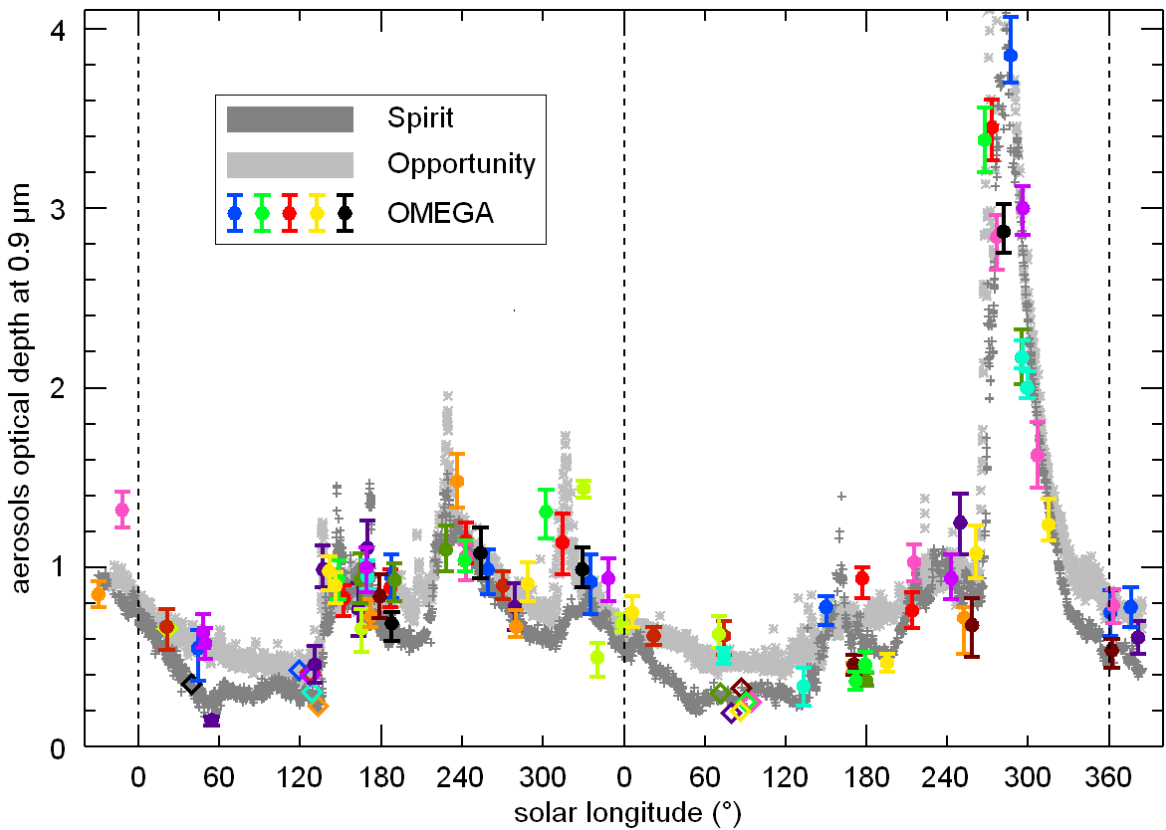

**Figure 9:** *Derived optical depths at 0.9 μm using OMEGA observations over dark surfaces (color points; relative error bars are indicated), plotted with PanCam/MERs nearly simultaneous measurements of the optical depth at 0.9 μm (Lemmon et al., 2004, 2006, and personal communication; stars: Opportunity; crosses: Spirit). OMEGA retrievals use the phase function of Ockert-Bell et al. (1997). Each color corresponds to a different surface region (see Figure 2 for the location). Diamonds indicate the assumed optical depth of the correction for the low optical depth spectrum of each sequence.*

### 4.3. Dust deposits at the surface

Assuming the dust deposition rate derived by Kinch et al. (2007) and the optical depths measured by the MERs (Lemmon et al., 2004, 2006 and personal communication), an amount of dust that corresponds to an optical depth of 0.5 (at 0.9 μm) should be deposited at the surface during the 90-sols long decay of the July 2007 global storm. Such dust deposits increase the reflectance factor of the surface from 0.1 to 0.2 (Kinch et al., 2007). The widespread occurrence of such transient bright dust deposits within the analyzed dark regions would therefore have induced a significant overestimation of the optical depth with our method. However, this is not observed (*Figure 9*). When taking into account the aerosol contribution and the specific photometric conditions, variations of the surface reflectance by more than a few % are not required in most cases to interpret the observed variations at different epochs (see *Figure 7* and *Figure 8*). This implies that the generic situation for dark regions during the 2004-2007 period is that either dust does not settle at the surface or dust is rapidly removed once deposited due to surface cleaning mechanisms (Sagan et al., 1972; Pleskot and Miner, 1981; Greeley et al., 2005; Rogers et al., 2007; Cantor et al., 2007). The lack of significant widespread dust deposits after the 2007 planet-encircling dust event contrasts with the large changes in surface albedo reported after the 2001 storm (Smith, 2004; Szwast et al., 2006). Different causes may explain this discrepancy. First, the 2007 dust storm occurred later in the season (between $L_S$ 265° and $L_S$ 325°, compared to $L_S$ 185° – 280°). The efficiency of cleaning processes occurring at dark regions may depend on season. For instance, the dust cover increase at Syrtis lasted until $L_S$ ~ 320° in 2001 (Szwast et al., 2006). This $L_S$ corresponds roughly to the end of the 2007 storm, which could explain why no persistent dust cover was detected after the 2007 storm. Secondly, the 2007 storm was shorter compared to that of 2001 (about 60° of $L_S$ versus 95°). Therefore, the amount of raised and settled dust was probably lower. Finally, we must notice that previous study of apparent albedo change do not always account for slight change in aerosols contribution. During clear atmospheric conditions ($\tau_{vis}$ ~ 0.4), a 0.1 change in optical depth from one year to another is enough to simultaneously increase the reflectance at 1 μm of dark regions by 0.01 while decreasing that of the brightest regions by a few 0.001. Such a change mimics a depletion of dust from bright regions and a deposition of dust above dark regions. T-IR optical depths are frequently used to assess similar atmospheric conditions and avoid this kind of mistakes (Szwast et al., 2006). However, a change of the visible optical depth can be undetectable in the thermal infrared if it is associated with a change of the mean particle size. As we will see in the next section, such particle size changes are observed after dust storm (*Figure 15*).

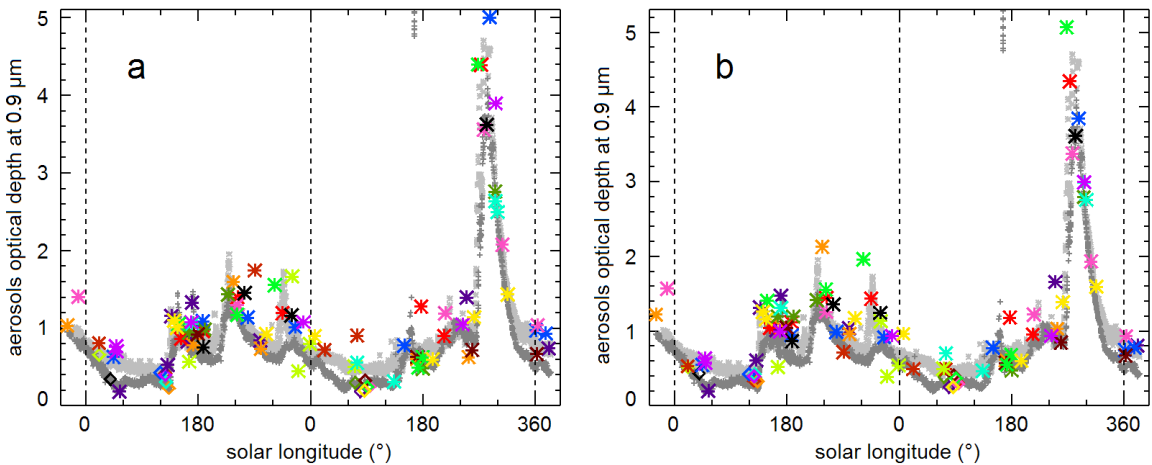

**Figure 10:** *Same as Figure 9 but for different assumptions: (a) We use the phase function of Tomasko et al. (1999). Optical depths are changed by a factor between 0.9 and 1.4 compared to the Ockert-Bell et al. function. (b) Retrieved optical depth with the Ockert-Bell phase function are scaled at an altitude of −1.7 km corresponding to the MER landing sites. We assume an exponential atmosphere and a constant scale height of 11.5 km to approximate well-mixed conditions. The elevation of the regions analyzed here varies between −4.5 and 3 km, which correspond to scaling factors that range from 0.7 to 1.3.*

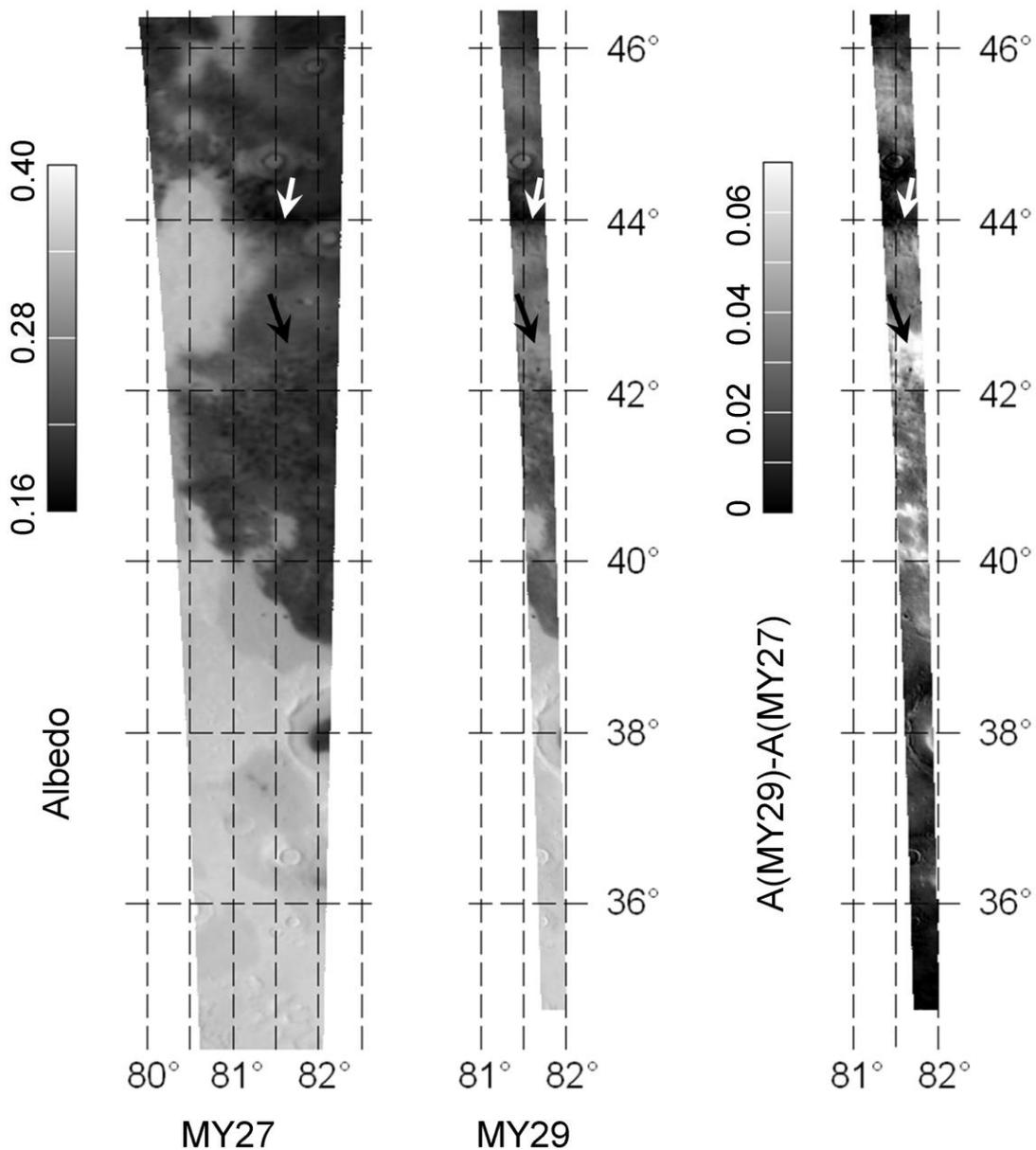

**Figure 11:** *(Left and middle) "Albedo maps" (RF at 1.08 μm) of the Nilosyrtis region that have undergone recent surface albedo changes (Geissler, 2005). (Left) $L_S$ 98° of MY27 (October 2004). (Middle) $L_S$ 20° of MY29 (January 2008), 30° of $L_S$ after the decay of a global dust storm. The two regions are observed with similar solar zenith angles (between 55° and 60°) and low optical depths according to MER measurements (approximately 0.4 for Ls 98° and 0.6 for Ls 20°). (Right) Albedo differences between MY29 and MY27. The albedo features are basically similar after and before the dust storm. Some places (notably that indicated by the black arrow) appear brighter, while others (white arrow) show nearly no changes of spectral properties(see Figure 12).*

We have searched for localized variations of surface reflectance in OMEGA observations by looking at albedo boundaries of dark regions that are known to undergo surface albedo changes (Geissler, 2005; Szwast et al., 2006) or to be covered by a dust coating (Rogers et al., 2007). Specific areas of interest were selected on the basis of the availability of OMEGA observations after the global dust storms of 2007. OMEGA observations of the Nilosyrtis region obtained in 2004 and 2008 with similar aerosols and lighting conditions are compared in *Figure 11*. Some areas show constant spectral properties, while some regions brighten (*Figure 12*). The observed brightening would imply an optical depth of 1.5 ± 0.15 with our method, while the optical depth is between 0.4 and 0.8 at that time. These increases in reflectance are not due simply to serendipitous occurrence of a local dust storms in the field-of-view during the observation, because no changes are observed in the depth of 2 μm $CO_2$ gas band (which is strongly correlated to the path length of photons in the atmosphere; Bibring et al., 1990). These differences occur on surfaces that manifest a strong negative spectral slope, which can be reproduced in laboratory experiments by a coating of dust (see e.g. Fischer and Pieters, 1993). The presence of such localized dust deposits is consistent with the observations and the analyzes of Sagan et al., (1972) which concluded that dust settles on the surface in specific areas linked with topographic features.

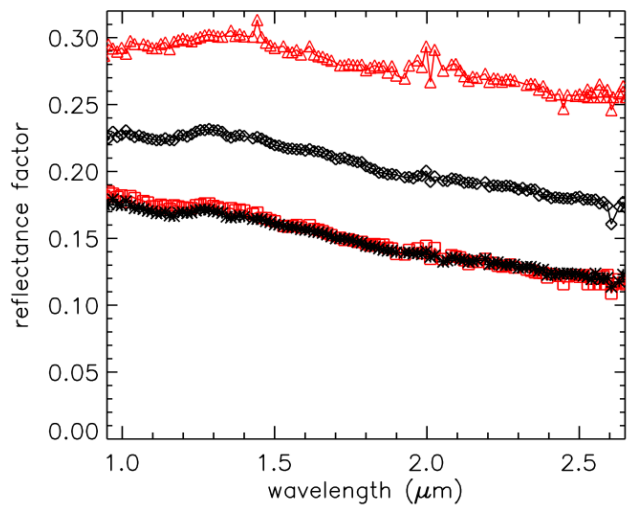

**Figure 12:** *Spectra corresponding to the regions highlighted by the arrows in Figure 11. Black symbols correspond to $L_S$ 98°, MY 27 and red symbols to $L_S$ 20°, MY 29. Most regions are the same before and after the dust storms (stars and squares: white arrow in Figure 11), but some regions undergo brightening (triangle and diamonds: black arrow in Figure 11). Spectra in that region show a strong decreasing slope even under low aerosols contribution, which could indicate a partial surface dust coating (Fischer and Pieters, 1993).*

### 4.4. Variations in the size of aerosols

The decrease in the cross section of aerosols with wavelength becomes steeper as the particle size decreases (Clancy et al., 2003). A relationship between the slope factor S and the mean particle size is derived by computing the scattering properties of aerosols with different effective radius $r_{eff}$ using the T-matrix code of Mishchenko and Travis (1998). We select the scattering properties of randomly oriented oblate cylinders (2:1 axial ratio) and a size distribution width $v_{eff}$=0.3, similarly to Wolff et al. (2006). The refractive indices are taken from Ockert-Bell et al. (1997). We use the factor $\omega Q_{ext} P(\theta)$ (the product of the single scattering albedo, the extinction cross section and the phase function at an angle $\Theta$) instead of the extinction cross section so as to account for the variations of the phase function with wavelength (*Figure 6*). $\omega Q_{ext} P(\theta)$ is proportional to the number of photons scattered once toward the zenith direction for an initial solar incidence angle $\pi - \theta$. This parameter is equivalent to the slope factor S when single scattering dominates.

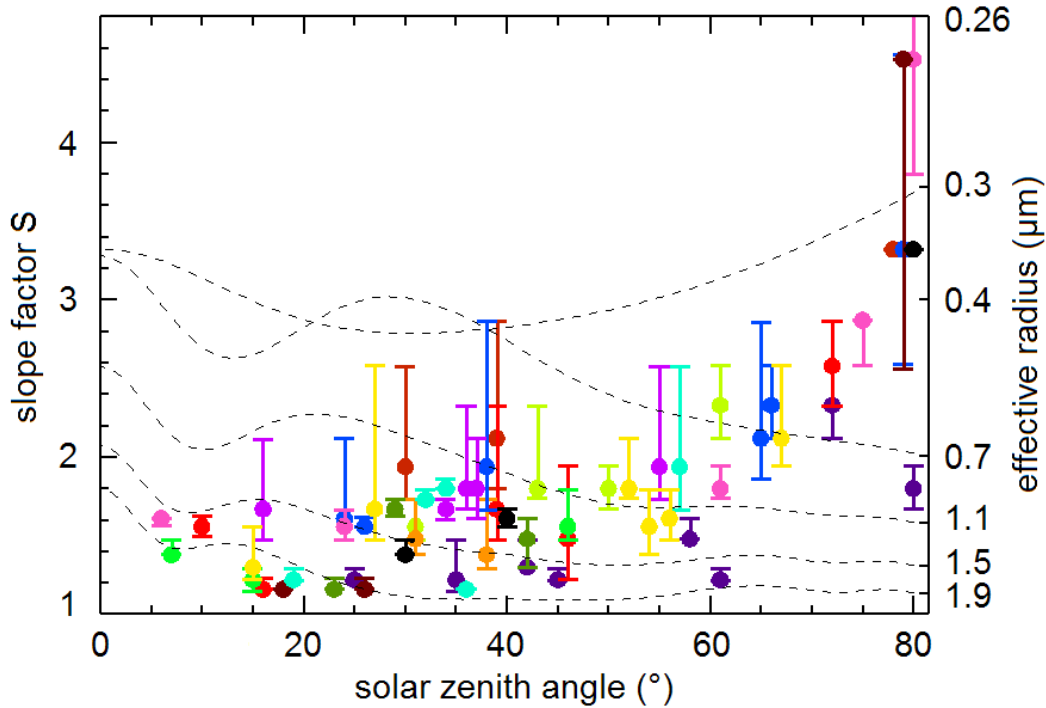

**Figure 13:** *Variations of the slope factor S as a function of the solar incidence angle (color points; relative error bars are indicated). The corresponding effective particle sizes are indicated with dotted lines (right y-axis). Particles vary with time within the 1–2 µm range for low to moderate incidence angle. Smaller mean particle sizes (<1 µm) are seen at high solar incidence angle (>60°). Observations with emergence angles greater than 40° are not considered, as well as points for which the error bar is greater than two.*

The variations of the aerosol size with solar zenith angle are indicated on *Figure 13*. Using the slope factor as an indicator of particle size, the effective size retrieved from observations obtained at low to moderate solar zenith angles varies with time in the 1 µm to 2 µm range. This result is consistent with that obtained by previous authors. Clancy et al. (2003) observed three different mean particle sizes on Mars with TES data: 1µm, 1.5µm and 1.8-2.5µm. Wolff and Clancy (2003) observed variations of $r_{eff}$ with time and space around 1.6 µm. Wolff et al. (2006) and Soderblom et al. (2008) observed an effective radius of aerosols that range between 1.2 µm and 1.8 µm at the MER landing sites. The effective particle size appears to decrease to values lower than 1 μm as the incidence angle goes past 60° (*Figure 13*). In fact, the slope factor derived from observations at extreme solar incidence (80°) implies quite small particle sizes (a few tenths of μm). We have simulated the altitude at which the first interaction with aerosols occurs as a function of the solar incidence angle with our Monte-Carlo model of radiative transfer (*Figure 14*). For a typical optical depth range of 0.5 – 1 (at 0.9 μm) and a scale height of 11.5 km that correspond to well-mixed conditions (Lemmon et al., 2004; Zazova et al., 2005), the mode of the altitude distribution increases from 0 (ground) at low to moderate solar incidence angle to 15-20 km for a solar incidence of 80°. Therefore, it may be reasonable to interpret the decrease of the mean particle size with solar incidence angle as a sensitivity of the OMEGA observations to a decreasing particle size with altitude. Several authors have already pointed out such a decrease of the particle size with altitude using ISM/Phobos and SPICAM/Mars Express data. Chassefière et al. (1992) retrieve a vertical gradient of 0.05 μm km$^{-1}$ below 25 km for the effective radius of dust particles. Korablev et al. (1993) found that the effective radius decreases from 1.6 μm near 15 km to 0.8 μm at 25 km. Montmessin et al. (2006) observe effective radius lower than 0.1 μm above 60 km and size larger than 0.3 μm below 50 km, and Rannou et al. (2006) observe dust particles that range between 0.01 and 0.1 μm above 20 km. This decrease in particle size with altitude is generally explained through gravitational settling (e.g., Montmessin et al., 2002). The mean altitude of interaction at high solar zenith angle depends on the optical depth (*Figure 14*): when the optical depth is very low (<0.2), most photons interact near the surface even at i=80°. This is consistent with the interpretation of large particles at high solar zenith angle and low optical depth ("purple" point at i=80° in *Figure 13*, which corresponds to an optical depth of 0.19 at L$_S$ 54°).

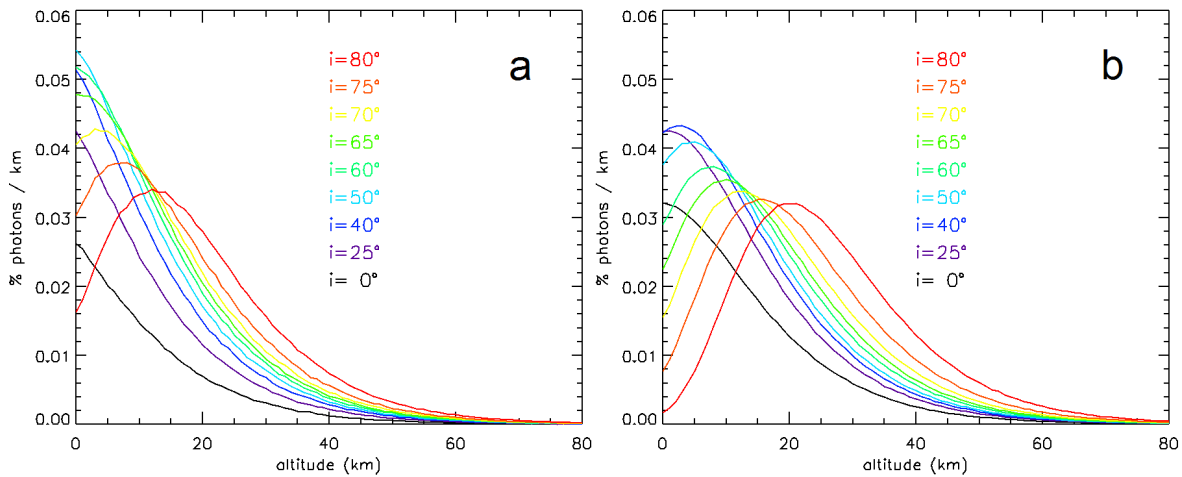

**Figure 14:** *Distribution of photons as a function of the altitude for the first interaction event (with dust aerosols) for different solar incidence angles. The dust optical depth varies with altitude with a scale height of 11.5 km. The total optical depth is 0.5 (a) and 1 (b). The altitude at which photons interact with aerosols increases with incidence angle, i.e., photons interact higher in the atmosphere at higher optical depth.*

Several authors have retrieved higher mean particle sizes during high opacity periods (Clancy et al., 2003; Wolff and Clancy, 2003; Wolff et al., 2006; Soderblom et al., 2008). Such behavior is predicted by numerical simulations (Murphy et al., 1993; Kahre et al., 2008) as larger particles lifted during storms settle faster than smaller particles. In *Figure 15*, we have plotted the variation of the particle size as a function of the optical depth for the period covering the maximum and the decay of the global dust storm of 2007 ($L_S$ 265° - $L_S$ 30°). Indeed, we observe a decrease of the particle size from 1.5 µm to 1.1 µm as the optical depth decreases after the dust storm, similarly to Wolff and Clancy (2003).

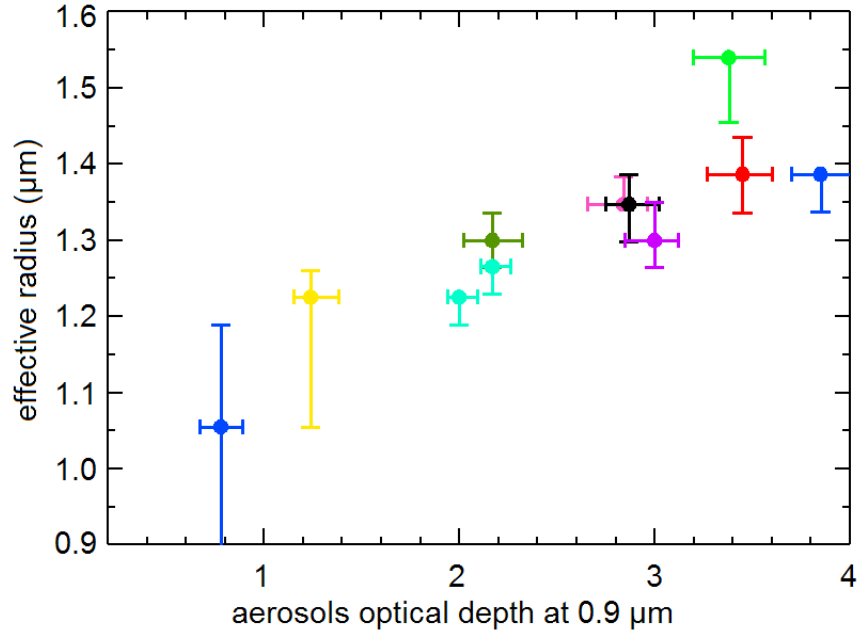

**Figure 15:** *Variations of the effective radius of aerosols as a function of the optical depth during and after the July 2007 dust storm (LS > 265°, MY28). We have selected observations with solar zenith angle <70°. Relative error bars are indicated (see Section 4.1 for details). The particle size decreases during the decay of the dust storm, which indicates that large particles settle faster than small particles.*

5.      **Conclusion**

We show that the observed variations of the apparent reflectance of dark surfaces in the OMEGA dataset from 2004 to 2007 can be reproduced by variations in the relative contribution of aerosols (due to different lighting conditions and column optical depths). The temporal variations of the optical depths derived from the observations of low to mid-latitudes dark regions (40°S - 40°N) are consistent with those determined at both MER sites. When combined with the similar evolutions at the two MER sites at longitudes 180° apart, our observations would seem to support the idea that the diffuse component of dust optical depth is relatively homogeneous over low to mid-latitudes and nearly independent of the longitude for a given epoch.

Widespread transient dust deposits are not observed over dark areas in the near-IR after the major planet encircling dust event of 2007. The decrease of the apparent reflectance of dark surfaces during the decay of dust storms is consistent with that expected from the decrease of the contribution of aerosols. This implies that either dust is not significantly deposited on most dark terrains or dust is rapidly removed by cleaning mechanisms. Localized surface deposits after the 2007 dust storms are observed at the boundaries between bright and dark regions.

The observed spectral characteristics of aerosols vary with time and photometric angles. Within the framework of our aerosol model, these variations are consistent with changes of the particle size of the aerosols. An effective radius between 1 µm and 2 µm is retrieved from OMEGA nadir observations at low to moderate solar incidence angles. This range corresponds to the variations with time of the mean particle size of the whole aerosols layer. The mean particle size of aerosols decreases during the decrease of the optical depth following the 2007 global dust storm. The apparent particle size seems to decrease significantly (< 1µm) for high solar incidence angles. Monte-Carlo radiative transfer simulations suggest that these variations are due to a decrease of the mean particle size with altitude in the bottom 30-40 kilometers of the atmosphere.

These results have implications for the retrieval of actual surface reflectance properties in the VIS-NIR spectral range by removing the contribution of aerosol scattering (Vincendon et al., 2007a, 2008; McGuire et al., 2008). The optical depths derived by PanCam on both Mars Exploration Rovers provide a useful first estimate of the optical depth for low to mid-latitudes regions, and can be used to evaluate the aerosol contributions for simultaneous observations of Mars in the near IR (OMEGA, CRISM) (Poulet et al., 2008). The implied variations of the particle size with time and altitude could significantly modify the spectral impact of aerosols on retrieved surface properties. Due to the non-repeatability of the Martian dust "cycle", a simultaneous measurement of the aerosols properties is clearly of interest. For CRISM on MRO, this is available as an Emission Phase Function that is acquired for each targeted observation (Murchie et al., 2007).

**Acknowledgment:**

The authors deeply thank M. T. Lemmon for making his optical depth retrievals available for our study. We would also like to thank Lori Fenton and an anonymous reviewer for useful comments and suggestions.